\documentclass[twocolumn,showpacs,preprintnumbers,amsmath,amssymb]{revtex4}

\usepackage{graphicx}
\usepackage{dcolumn}
\usepackage{bm}

\begin{document}
\title{Evidence for a new magnetic field scale in CeCoIn$_{5}$}
\author{I. Sheikin$^{1}$, H. Jin$^{2}$,  R. Bel$^{2}$, K. Behnia$^{2}$, C. Proust$^{3}$,
 J. Flouquet$^{4}$, Y. Matsuda$^{5}$, D. Aoki$^{4}$ and Y. \={O}nuki$^{6}$}

\affiliation{
(1)Grenoble High Magnetic Field Laboratory (CNRS), BP 166, 38042 Grenoble, France \\
(2)Laboratoire de Physique Quantique (CNRS), ESPCI, 10
Rue de Vauquelin, 75231 Paris, France \\
(3)Laboratoire National des Champs Magn\'etiques Puls\'es (CNRS),
BP 4245, 31432 Toulouse, France \\
(4)DRFMC/SPSMS,  Commissariat \`a l'Energie Atomique, F-38042
Grenoble, France\\
(5)Department of Physics, University of Kyoto, Kyoto 608-8502, Japan\\
(6)Graduate School of Science, Osaka University, Toyonaka, Osaka 560-0043, Japan}

\date{\today}

\begin{abstract}
The Nernst coefficient of CeCoIn$_{5}$ displays a distinct anomaly at $H_{\rm_K} \sim$ 23 T. This feature is
reminiscent of what is observed at 7.8 T in CeRu$_{2}$Si$_{2}$, a well-established case of metamagnetic transition. New
frequencies are observed in de Haas-van Alphen oscillations when the field exceeds 23 T, which may indicate a
modification of the Fermi surface at this field.
\end{abstract}

\pacs{75.30.Kz,  73.43.Nq, 71.27.+a}

\maketitle

Heavy-Fermion (HF) compounds \cite{flouquet} display a dazzling variety of physical phenomena which still lack a
satisfactory general picture. The "non-Fermi-liquid" behavior emerging in the vicinity of a magnetic quantum critical
point (QCP), associated with a continuous (i.e. second-order) phase transition at zero temperature, has recently
attracted much attention \cite{stewart}. The case of the HF superconductor CeCoIn$_{5}$ \cite{petrovic} is intriguing.
Magnetic field alters the normal-state properties of this system in a subtle way. In zero field, the system displays
neither a $T^2$ resistivity nor a $T$-linear specific heat (standard features of a Landau Fermi liquid) down to the
superconducting transition. When superconductivity is destroyed by the application of pressure \cite{sidorov} or
magnetic field \cite{paglione,bianchi}, the Fermi-liquid state is restored. In the latter case, the field-tuned QCP
identified in this way is pinned to the upper critical field, H$_{c2}$ of the \emph{superconducting} transition
\cite{bauer,ronning}. This is unexpected, not only because quantum criticality is often associated with the destruction
of a magnetic order, but also because the superconducting transition becomes first order at very low
temperatures\cite{bianchi2}. The possible existence of a magnetic order with a field scale close to H$_{c2}$ and
accidentally hidden by superconductivity, has been speculated \cite{bianchi} but lacks direct experimental support.

Comparing CeCoIn$_{5}$ with the well-documented case of CeRu$_{2}$Si$_{2}$ \cite{haen,flouquet2} is instructive. In the
latter system a metamagnetic transition occurs at $H_m$ = 7.8 T: the magnetization jumps from 0.6 $\mu_{\rm B}$ to 1.2
$\mu_{\rm B}$ in a narrow yet finite window( $\Delta H_m$ = 0.04 T in the $T = 0$ limit). The passage from an
antiferromagnetically (AF) correlated system below $H_m$ to a polarized state dominated by local fluctuations above is
accompanied by a sharp enhancement in the quasi-particle mass in the vicinity of $H_m$ which is thus akin to a
field-tuned QCP. A sudden change of the Fermi surface (FS) topology across the metamagnetic transition has been
established by de Hass-van Alphen (dHvA) effect studies \cite{haoki}, where new frequencies were detected above $H_m$.

In this letter, we report on two sets of experimental studies which indicate that the effect of the magnetic field on
the normal-state properties of these two systems share some common features. They point to the existence of another
field scale in CeCoIn$_{5}$ which has not been previously identified. By measuring the Nernst coefficient and studying
quantum oscillations, we find compelling evidence that close to $H_{\rm_K} \sim$ 23 T, the FS is modified. Therefore,
in the $T = 0$ limit, CeCoIn$_{5}$ appears to display at least two distinct field scales.

Single crystals of CeCoIn$_{5}$ were grown using a self-flux method. Thermoelectric coefficients were measured using a
one-heater-two-thermometer set-up. dHvA measurements were done using a torque cantilever magnetometer. The magnetometer
was mounted in a top-loading dilution refrigerator equipped with a low-temperature rotation stage.

\begin{figure}
\includegraphics{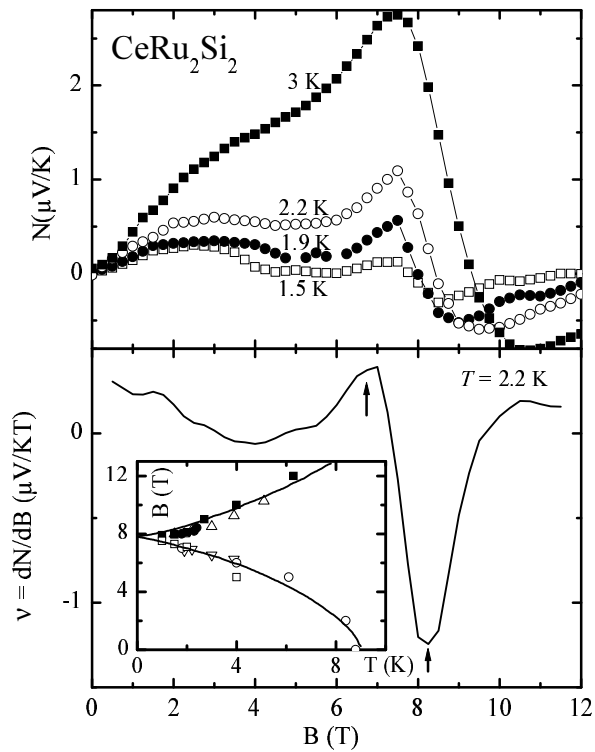}
\caption{\label{fig1}Upper panel: field-dependence of the Nernst signal in CeRu$_{2}$Si$_{2}$ for selected
temperatures. Lower panel: the Nernst coefficient, defined as the derivative of the signal at $T = 2.2$ K. The arrows
show the position of a maximum and a minimum close to the metamagnetic transition field. The inset shows the position
of these anomalies (triangles) in the $H-T$ plane compared to those detected by previous studies of thermal expansion
(solid and empty circles \cite{paulsen}) and specific heat (solid and empty squares \cite{yaoki,meulen}).}
\end{figure}

We begin by presenting the field-dependence of the Nernst coefficient in CeRu$_{2}$Si$_{2}$, which demonstrates the
remarkable sensitivity of this probe. Fig. 1 shows the field-dependence of the Nernst signal
(N=$\frac{E_{y}}{\nabla_{x}T}$) in CeRu$_{2}$Si$_{2}$. As seen in the upper panel of the figure, $N$ abruptly changes
sign around the metamagnetic transition field, $H_{\rm m}=7.8$ T. Besides this striking feature, the field-dependence
of the Nernst signal presents additional structure. The lower panel of the same figure shows the field-dependence of
the dynamic Nernst coefficient, $\nu=\frac{\partial N}{\partial B}$ at 2.2 K. It presents two anomalies just below and
above $H_{\rm m}$: a sharp minimum at $\sim$ 8.2 T and a smaller maximum at 7 T. The inset of the figure shows the
temperature-dependence of these two anomalies which closely follow the lines of the pseudo-phase diagram of
CeRu$_{2}$Si$_{2}$. These are crossover lines which represent anomalies detected by specific heat \cite{meulen,yaoki}
and thermal expansion \cite{paulsen} measurements.

The case of CeRu$_{2}$Si$_{2}$ shows how sensitively the Nernst signal probes metamagnetism. This is presumably due to
its intimate relationship with the energy dependence of the scattering rate (the so-called Mott formula:
$\nu=\frac{\pi^{2}k_{B}^{2}T}{3m}(\frac{\partial\tau}{\partial\epsilon})|_{\epsilon=\mu}$). With this in mind, let us
focus on the case of CeCoIn$_{5}$. The first study of the Nernst effect in this compound found a very large zero-field
Nernst coefficient emerging below $T^{*}\sim 20K$ \cite{bel}. Below this temperature, resistivity is linear in
temperature \cite{petrovic,paglione,nakajima}, the Hall coefficient is anomalously large \cite{nakajima,hundley}, the
thermoelectric power is anomalously small \cite{bel} and the electronic specific heat rises rapidly with decreasing
temperature \cite{petrovic,bianchi}. All these anomalous properties gradually disappear when a magnetic field is
applied. In particular, the giant Nernst effect also gradually fades away in presence of a moderate magnetic field
\cite{bel}.

The field-dependence of the Nernst coefficient in the $12 - 28$ T field range, reported here for the first time,
reveals new features emerging at still higher magnetic fields. The $H-T$ (pseudo-)phase diagram of the system is
apparently more complicated than previously suggested, and the field associated with the emergence of the Fermi-liquid
close to H$_{c2}(0)$ ($\sim $ 5T) is not the only relevant field scale for CeCoIn$_{5}$.

\begin{figure}
\includegraphics{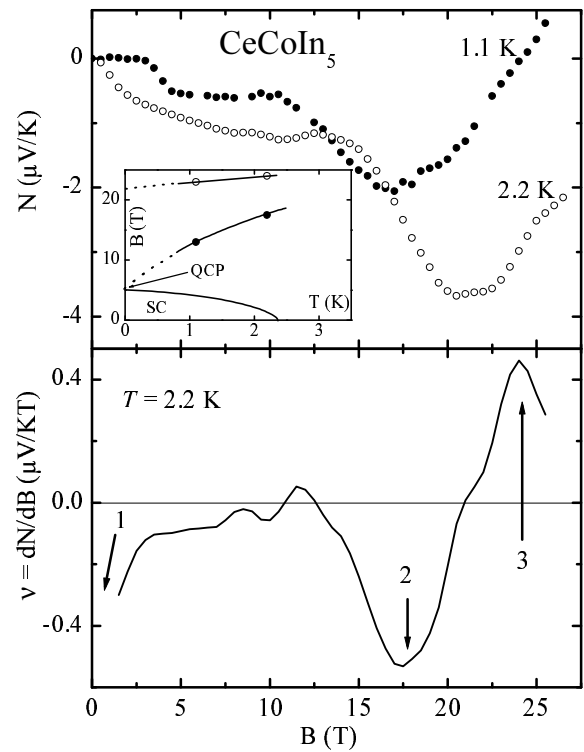}
\caption{\label{fig2}Upper panel: field-dependence of the Nernst signal in CeCoIn$_{5}$ for two temperatures. Lower
panel: the Nernst coefficient for 2.2 K obtained by taking the derivative of the data presented in the upper panel.
Arrows identify three anomalies in the 2.2 K curve.  A sketch of the $H-T$ phase diagram based on these anomalies is
shown in the inset of the upper panel.}
\end{figure}

Fig. 2 presents the field-dependence of the Nernst signal, $N$, and its dynamic derivative, $\nu$, in a magnetic field
along the $c$-axis and up to 28 T. At the onset of superconductivity ($T = 2.2$ K), $N$ rises rapidly as a function of
field at low fields. Thus, its derivative, $\nu$ is large (anomaly no.1). With the application of a moderate magnetic
field, the Nernst signal saturates and, therefore, $\nu$ becomes very small. This is the behavior  previously detected
and identified as the signature of a field-induced Fermi-liquid state. However, above 15 T, $N$ increases suddenly
again, reaches rapidly a maximum and then decreases. The field-dependence of $N$ at $T = 1.1$ K repeats the same scheme
with all field-scales shifted to lower values. The two anomalies of opposite signs revealed in $\nu(B)$ (marked no.2
and no.3 in the figure) are reminiscent of what was observed in the case of CeRu$_{2}$Si$_{2}$.

This indicates the presence of two distinct field scales in CeCoIn$_{5}$. However, contrary to the case of
CeRu$_{2}$Si$_{2}$, they do not tend to merge in the zero-temperature limit. The lower-field anomaly (no.2) lies close
to the line already identified by the resistivity measurements \cite{paglione}. At $T = 0$, this line ends up very
close to the superconducting upper critical field ($\sim$ 5 T for this field orientation).  The high-field anomaly (no.
3) identifies another (almost horizontal) line in the $H-T$ plane and yields a second field scale ($\sim$ 23 T).  A
sketch of the (pseudo-)phase diagram of CeCoIn$_5$ is shown in the inset of Fig. 2.

\begin{figure}
\includegraphics{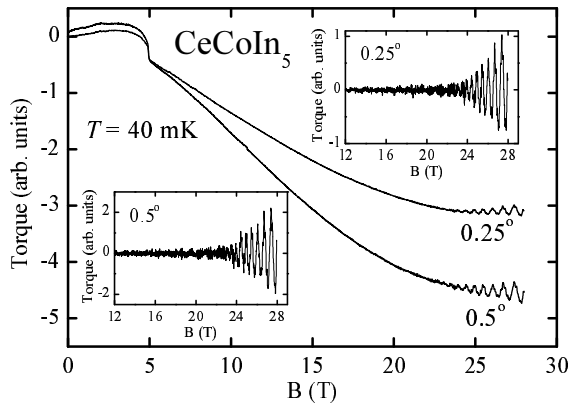}
\caption{\label{fig3} Field-dependence of the magnetic torque in CeCoIn$_{5}$ observed at 40 mK for two orientations of
the magnetic field close to the $c$-axis. The insets show the high field oscillatory torque signal after subtracting
the background.}
\end{figure}

After detecting the Nernst effect anomalies, we performed high-resolution dHvA measurements at high fields. They
provide another evidence for the existence of another field scale at $H \sim$ 23 T in CeCoIn$_{5}$. Fig. 3 shows the
torque signal at $T = 40$ mK as a function of magnetic field applied close to the $c$-axis. Clear anomalies observed at
around 5 T correspond to the suppression of superconductivity. There are no other remarkable anomalies at higher field,
in particular similar to that observed at 9 T in CePd$_2$Si$_2$, where a metamagnetic transition was established by
torque measurements \cite{sheikin}. However, a sudden emergence of a new dHvA frequency is clearly detected above $\sim
23$ T and becomes evident after subtracting the background (insets of Fig. 3). Remarkably, the amplitude of the new
frequency is so strong that it dominates all the other dHvA oscillations above 23 T at very small inclinations from the
$c$-axis.

\begin{figure}
\includegraphics{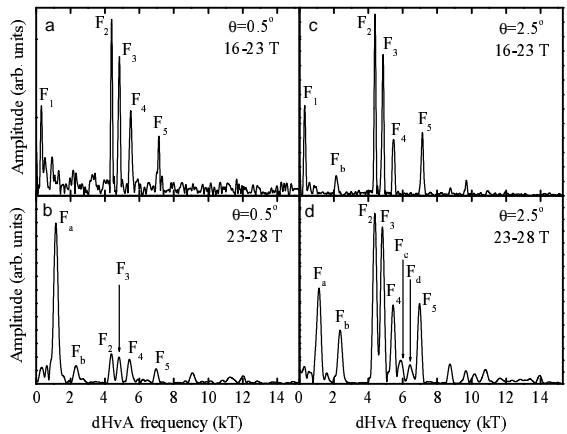}
\caption{\label{fig4}Fourier spectra of the dHvA oscillations below (a and c) and above (b and d) 23 T are compared for
magnetic field applied at 0.5$^\circ$ (a and b) and 2.5$^\circ$ (c and d) from the $c$-axis. Only the fundamental
frequencies are denoted by letter. The frequencies that are present both below and above 23 T and were observed in the
previous measurements are denoted by $F$ with a numeric index, while the new frequencies that appear above 23 T only
are denoted by by $F$ with a letter index.}
\end{figure}

\begin{table}
\caption{\label{table1}dHvA frequencies and corresponding effective masses observed below and above $H_{\rm_K} \sim$ 23
T.}
\begin{ruledtabular}
\begin{tabular}{ccccc}
&\multicolumn{2}{c}{$16-23$ T}&\multicolumn{2}{c}{$23-28$ T}\\
&$F$ (kT)&$m^* (m_0)$&$F$ (kT)&$m^* (m_0)$\\
\hline
$\theta = 0.5^\circ$&&&&\\
$F_1$& 0.31 & 11.8 & 0.32 &\\
$F_a$& & & 1.15 & 34.5 \\
$F_b$& & & 2.34 & 65.7 \\
$F_2$& 4.39 & 7.4 & 4.37 & 6.0 \\
$F_3$& 4.84 & 12.8 & 4.81 & 14.5 \\
$F_4$& 5.48 & 18.4 & 5.42 & 13.2 \\
$F_5$& 7.14 & 24.2 & 6.97 & 26.0 \\
\hline
$\theta = 2.5^\circ$&&&&\\
$F_1$& 0.30 & 10.0 & 0.29 & 4.41 \\
$F_a$& & & 1.13 & 33.4 \\
$F_b$& 2.12 & & 2.36 & 48.8 \\
$F_2$& 4.39 & 6.8 & 4.38 & 4.3 \\
$F_3$& 4.86 & 11.2 & 4.81 & 15.2 \\
$F_4$& 5.47 & 14.2 & 5.44 & 11.0 \\
$F_c$& & & 5.89 & 18.3 \\
$F_d$& & & 6.42 & 23.6 \\
$F_5$& 7.15 & 31.3 & 6.98 & 27.0 \\
\end{tabular}
\end{ruledtabular}
\end{table}

The field-induced emergence of new dHvA frequencies becomes evident when comparing the Fourier spectra of the dHvA
oscillations below and above 23 T. Such a comparison is shown in Fig. 4 for magnetic field applied at 0.5$^\circ$ and
2.5$^\circ$ from the $c$-axis, the two orientations for which the effective masses were measured. The dHvA frequencies
and corresponding effective masses for these two orientations are given in Table I. All the fundamental frequencies
observed both below and above 23 T are marked in the figure. The frequencies $F_i, i=1...5$ are observed both below and
above 23 T and are in good agreement with those found in the previous dHvA studies performed at lower fields
\cite{settai,hall,McCollam}. It was shown \cite{settai} that all these frequencies correspond to quasi-two-dimensional
FS that are well accounted for by the itinerant $f$-electron band structure calculations.

For the magnetic field orientation at 0.5$^\circ$ from the $c$-axis, two new frequencies $F_a$ and $F_b$ appear in the
oscillatory spectrum above 23 T (Fig. 4 b). Neither of them was observed in the previous lower field dHvA studies. The
corresponding effective masses of the new frequencies are quite high, being of the order of 35 $m_0$ and 65 $m_0$. The
frequency $F_a$ might correspond to one of the closed orbits of the 15-electron band of the band structure calculations
\cite{settai}. In this case, however, there is no reason for this frequency not being observed at lower field. Even
taking into account its high effective mass and its presumable field dependence it should have been observed at lower
fields since its effective mass is considerably smaller than that of $F_b$, while the amplitude is much stronger. The
other frequency, $F_b$, can not be reconciled with any orbit from the theoretical calculations.

For the magnetic field at 2.5$^\circ$ to the $c$-axis (Fig. 4 c and d), $F_a$ and $F_b$ with high effective masses are
also present above 23 T. Furthermore, two more higher frequencies, $F_c$ and $F_d$ emerge above 23 T for this
orientation (Fig. 4 d). Like $F_a$, these two frequencies were not observed either below 23 T or in the previous
studies. The effective masses corresponding to these frequencies are also quite enhanced being about 18 $m_0$ and 24
$m_0$ respectively.

The frequencies $F_a$ and $F_b$ exist only over a small angular range close to the $c$-axis. Therefore, they might be
due to a magnetic breakdown. This is not the case of the frequencies $F_c$ and $F_d$ which survive over a wide angular
range between the [001] and [100] directions. Their emergence above 23T seems to imply an important modification of the
FS topology. Since these new frequencies are associated with large effective masses, the drastic change observed in the
Nernst signal at 23T would be a natural consequence of such a modification.

Thus, two independent sets of evidence points to the existence of a new field scale in CeCoIn$_{5}$ at 23 T:  A drastic
change in the magnitude and sign of the Nernst signal and the appearance of the new frequencies in dHvA spectrum.  Both
these features have been observed in CeRu$_{2}$Si$_{2}$ as experimental signatures of the metamagnetic transition. Let
us also note that a temperature scale of $\sim$ 20 K, (comparable to the energy associated with a field of 23 T) marks
most of the zero-field properties of the heavy-electron fluid in CeCoIn$_{5}$.  However, what occurs in the case of
CeCoIn$_{5}$ does not appear as a metamagnetic transition \cite{shishido} (defined as a jump in the magnetization in a
narrow field window). Let us discuss the possible origins of this feature below.

In HF systems, the interplay between magnetic intersite interactions with a characteristic energy scale $E_m$ and local
Kondo fluctuations with a characteristic energy $k_{\rm B} T_{\rm K}$ is changed with the application of a magnetic
field. When the associated Zeeman energy becomes comparable with one of the characteristic energy scales, the balance
between these two interactions is modified. This defines two characteristic field scales, $H_m$ and $H_{\rm K}$, given
by $g \mu H_m \simeq E_m$ and $g \mu H_{\rm K} \simeq k_{\rm B} T_{\rm K}$. Furthermore, an external magnetic field
adds a ferromagnetic (F) component to the existing AF one. The field redistribution among different AF, F and Kondo
components depends on the type of the local anisotropy (Ising, planar or Heisenberg character) and the lattice
deformation produced by magnetostriction. In this context, an important difference between CeRu$_{2}$Si$_{2}$ and
CeCoIn$_{5}$ is that the susceptibility in the latter compound is much less anisotropic. CeRu$_{2}$Si$_{2}$, with a
magnetic anisotropy of 15 compared to 2 in CeCoIn$_{5}$ is much closer to an Ising-like system. In the latter compound,
magnetic field seems to induce a QCP, i.e. the N\'{e}el temperature vanishes at $H_{c2}(0)$ \cite{paglione,bianchi}.
Due to the weak magnetic anisotropy of CeCoIn$_{5}$ $H_m$ should vanish at the field induced QCP \cite{flouquet} as was
observed in YbRh$_{2}$Si$_{2}$ \cite{tokiwa} and CeNi$_{2}$Ge$_{2}$ \cite{gegenwart}. For CeCoIn$_{5}$, this may
explain the existence of a large field domain between $H_{c2}$ and $H_{\rm K}$ where AF and F correlations compete.

The possible occurrence of a metamagnetic transition in the CeMIn$_{5}$ family (M = Co, Rh or In) has been a subject of
recent debate \cite{takeuchi,kim,palm,capan}. Several studies suggest a metamagnetic transition in CeIrIn$_{5}$ in the
$25 - 42$ T field range \cite{takeuchi,kim,palm,capan}.  On the other hand, the modification of the FS topology
observed here may be closely related to the drastic change of the FS recently detected in CeRhIn$_5$ at $P =$ 2.4 GPa,
just above its critical pressure $P_c =$ 2.35 GPa \cite{shishido1}. The opposite effects on the FS produced by pressure
and magnetic field are in good agreement with the opposite actions on the volume (contraction and expansion
respectively).

In conclusion, we have demonstrated the existence of another field scale in CeCoIn$_5$ besides $H_{\rm c2}$. The new
characteristic field $H_{\rm K} \sim 23$ T is marked by an anomaly in the Nernst signal reminiscent of that observed at
the metamagnetic field of CeRu$_2$Si$_2$. The characteristic field is also associated with the appearance of new
frequencies in dHvA oscillations, like in CeRu$_2$Si$_2$. However, contrary to the latter system, there appear to be
two distinct field scales widely separated in the T = 0 limit.


\end{document}